\documentclass[twocolumn,floats,floatfix,superscriptaddress,prl,nofootinbib]{revtex4}
\usepackage{float}
\usepackage{graphicx}
\usepackage{amsmath}
\usepackage{amsfonts}
\usepackage[utf8]{inputenc}

\usepackage{color}

\begin{document}

\title{Algorithmic Lattice Kirigami: A Route to Pluripotent Materials}
\author{Daniel M. Sussman}\thanks{DMS and YC contributed equally to this work}
\affiliation{Department of Physics and Astronomy, University of Pennsylvania, 209 South 33rd Street, Philadelphia, Pennsylvania 19104, USA}
\author{Yigil Cho}
\affiliation{Department of Physics and Astronomy, University of Pennsylvania, 209 South 33rd Street, Philadelphia, Pennsylvania 19104, USA}
\affiliation{Department of Materials Science and Engineering, University of Pennsylvania, 3231 Walnut Street, Philadelphia, Pennsylvania 19104, USA}
\author{Toen Castle}
\affiliation{Department of Physics and Astronomy, University of Pennsylvania, 209 South 33rd Street, Philadelphia, Pennsylvania 19104, USA}
\author{Xingting Gong}
\affiliation{Department of Physics and Astronomy, University of Pennsylvania, 209 South 33rd Street, Philadelphia, Pennsylvania 19104, USA}
\author{Euiyeon Jung}
\affiliation{Department of Materials Science and Engineering, University of Pennsylvania, 3231 Walnut Street, Philadelphia, Pennsylvania 19104, USA}
\author{Shu Yang}
\affiliation{Department of Materials Science and Engineering, University of Pennsylvania, 3231 Walnut Street, Philadelphia, Pennsylvania 19104, USA}
\author{Randall D. Kamien}\email{kamien@upenn.edu}
\affiliation{Department of Physics and Astronomy, University of Pennsylvania, 209 South 33rd Street, Philadelphia, Pennsylvania 19104, USA}
\date{\today}

\begin{abstract}
We use a regular arrangement of {\sl kirigami} elements to demonstrate an inverse design paradigm for folding a flat surface into complex target configurations. We first present a scheme using arrays of  disclination defect pairs on the dual to the honeycomb lattice; by arranging these defect pairs properly with respect to each other and choosing an appropriate fold pattern a target stepped surface can be designed. We then present a more general method that specifies a fixed lattice of {\sl kirigami} cuts to be performed on a flat sheet. This single ``pluripotent'' lattice of cuts permits a wide variety of target surfaces to be programmed into the sheet by changing the folding directions.
\end{abstract}

\maketitle

\section{Introduction}
Reduced-dimensionality objects can be both light yet extremely strong, as in a geodesic dome or the skeleton of a diatom \cite{diatom}. Design of such structures couples scale-independent geometry and topology with material properties to create a variety of structures. In soft gel sheets, programmed inhomogeneous swelling and stretching can generate tunable three-dimensional shapes that can potentially serve as compliant mechanisms \cite{sharon}. In more rigid systems, {\sl origami} applies a prescribed sequence of folds to a flat sheet and returns a strong, lightweight, and flexible three-dimensional structure. A great deal of effort has gone into exploring the breadth of attainable {\sl origami} surfaces starting from nearly unstretchable sheets \cite{miura,maha,torus}, including recent work on designing mechanical metamaterials \cite{ori_metamat} and self-folding {\sl origami} structures \cite{self_origami}.

In {\sl origami}, the ``inverse problem'' of prescribing a set of folds to achieve a target structure has been algorithmically solved. For instance the circle/river packing method can be used to create the 2D ``base'' of the final product which is then augmented with extra folds into the desired 3D shape \cite{lang}. A different method wraps the initial flat sheet onto a polygonally-tiled target surface, using ``tucking molecules'' to hide the excess material within the final shape, and thus creates curvature \cite{tachi3d}. Further, software has been developed by the same author to form developable, irregularly corrugated target configurations by modifying {\sl miura-ori} style patterns \cite{tachi2d}.

There are, nevertheless, certain limits and constraints in the use of {\sl origami} to design generic structures. These problems center around the potential complexity of the initial fold pattern, and the subsequent greatly magnified complexity of the required sequence of folds along this pattern.  For instance, the fold patterns specified by the circle/river packing algorithms return a pattern whose folded state matches a target surface, but there is no guarantee that any subset of creases can be folded; complex models of this type are typically pre-creased and then folded in a very specific and repetitive sequence. The polygonally-tiled surfaces created by the tucking-molecule method are the product of intricately interlocking crimp folds hidden away beneath the surface. Both of these features reflect the fact that typically complex origami designs do not permit a monotonic, continuous folding motion from the planar state to the desired target state. These extremely delicate fold patterns present an obvious challenge to the goal of designing self-assembling {\sl origami} structures, where in many cases extremely fine control of the fold ordering may be required. Additionally, these fold patterns often waste much of the surface to create effective areas of Gaussian curvature, using extremely intricate folds, wedges, and pleats to effectively ``remove'' material, hiding it beneath the visible surface.

Recently we introduced lattice {\sl kirigami} methods to this design problem \cite{Castle2014}; inspired by the work of Sadoc, Rivier, and Charvolin on phyllotaxis \cite{sadoc1,sadoc,sadoc2} we supplemented the folds of {\sl origami} with a limited set of cutting and re-gluing moves taking place on a honeycomb lattice \cite{Castle2014}. The essence of our {\sl kirigami} constructions is that after making the prescribed cuts and identifying edges we have a surface with localized points of Gaussian curvature, which cause the surface to buckle into a three-dimensional configuration. Associated folding then precisely defines the shape of the surface, corralling the points of Gaussian curvature into useful cues that direct the shape of the final structure.

The restrictions imposed on {\sl kirigami} by the honeycomb lattice and its triangular dual lattice led to a manageable set of allowed motifs: 2-4 disclination pairs on the honeycomb lattice and $\tilde 5$-$\tilde 7$ disclination pairs on the dual lattice (dual lattice constructs are denoted by tildes) that could be connected to other disclination pairs with cancelling Burgers vectors by paths with glide and (sometimes partial) climb geometries. Another allowed {\sl kirigami} motif is the ``sixon,'' a defect formed by completely removing one hexagon from the honeycomb lattice and reconnecting the cut edges of the remaining surface in one of several degenerate ways. One configuration of the sixon identifies pairs of adjacent edges so that three triangular plateaus emanate from a central point; another configuration leads to matching 2-4 pairs in a ``pop-up/pop-down'' configuration \cite{Castle2014}. 

In this paper we show that arrangements of these {\sl kirigami} motifs enables a remarkably versatile ability to design patterns whose cut-and-folded state matches a desired target surface, allowing us to algorithmically solve the inverse design problem for a particular class of surfaces with a specified maximum gradient. 

Of particular interest in the potential for self-assembling designs, the {\sl kirigami} design paradigms we propose are simple -- to the extent of being easily designed by hand -- and robust to the ordering and relative rates of edge folding. In what follows we will first describe a simple solution to the inverse-design problem of creating a stepped surface with assigned heights at each point by arranging paired $\tilde 5$-$\tilde 7$ pairs along the lattice vectors of a larger superimposed honeycomb lattice. The excised patches define the location of the points of Gaussian curvature and the designation of mountain and valley folds defines the height. We then present a second method using the same principle in a more refined manner to produce a reconfigurable surface. By invoking the degeneracy and flexibility of the sixon structure we can design a pluripotent lattice of sixons that can be folded into a dizzying array of target shapes by selectively assigning their mountain and valley folds.

\section{Design using $\tilde 5$-$\tilde 7$ elements}

\begin{figure}
\centerline{\includegraphics[width=0.5\textwidth]{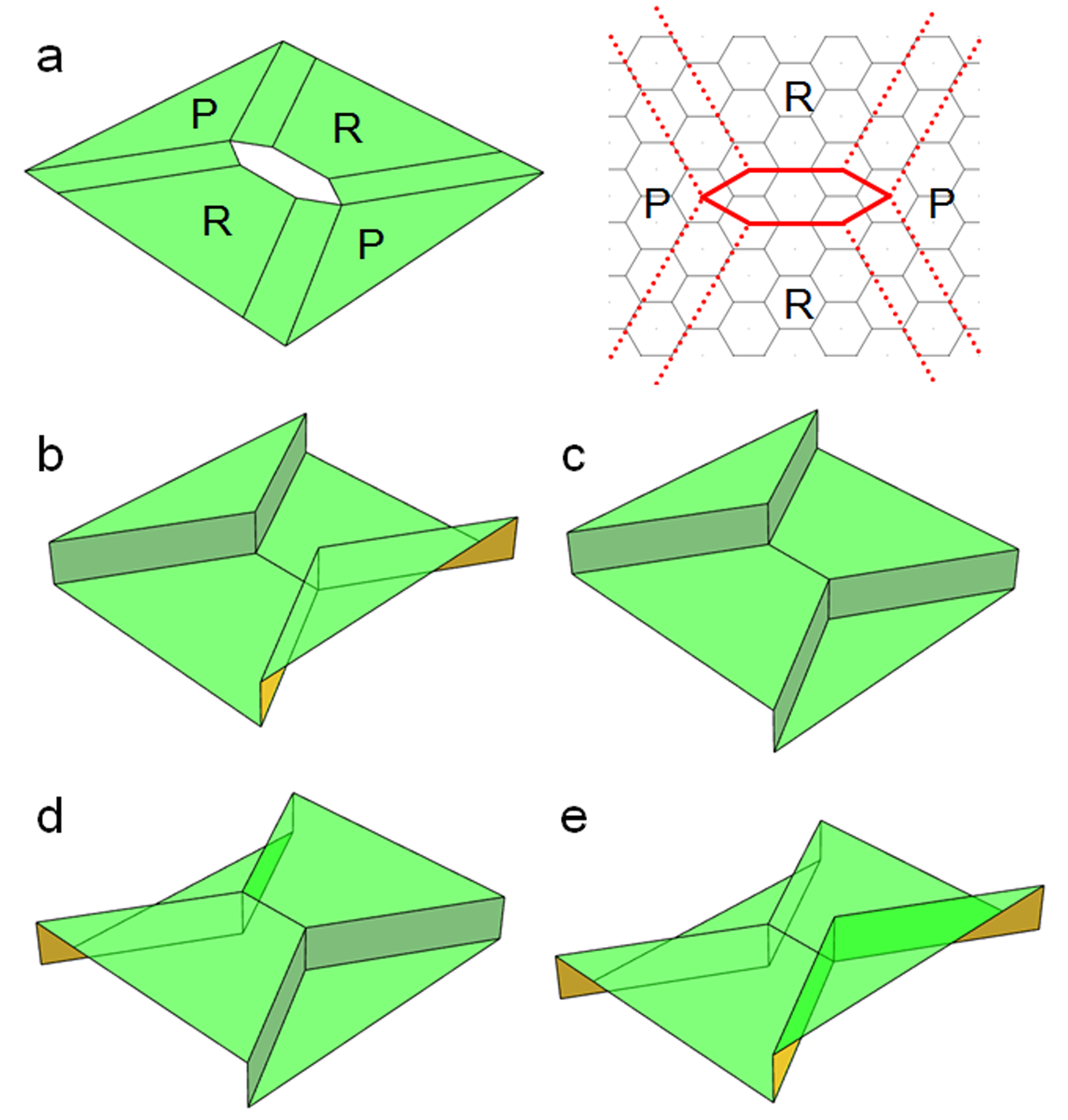}}
\caption{\label{fig:5_7_iso}
Construction and configuration of the fundamental $\tilde 5$-$\tilde 7$ climb pair {\sl kirigami} element. (a) The cut surface in its unfolded state with and without the underlying honeycomb lattice. Since they share an edge after assembly regions marked ``R'' must be at the same height in the folded configuration. Regions marked ``P'' can be either at the same height or differ in height by two. (b) -- (e) The four allowed folding configurations of the $\tilde 5$-$\tilde 7$ climb pair element.
}
\end{figure}

Figure \ref{fig:5_7_iso} depicts a single $\tilde 5$-$\tilde 7$ climb pair element, where an extended hexagon is excised from the sheet, its edges are identified, and mountain and valley folds perpendicular to the short sides of the excised hexagon are applied. The two regions labeled ``P'' can independently pop-up or pop-down relative to their initial configuration, depending on the fold choices. In contrast, the regions labeled ``R'' share an edge in the folded configuration, and therefore must be at the same height. Thus, an isolated $\tilde 5$-$\tilde 7$ climb pair element has four allowed configurations if we do not allow rotations in $\mathbb{R}^3$.

A workable design paradigm requires combining many of these kirigami elements together, and in general one of the challenges of kirigami design is to understand the allowed relative configurations of kirigami elements: Important progress can be made by realizing that a collection of $\tilde 5$-$\tilde 7$ climb pair elements can be placed on the edges of a larger honeycomb ``super-lattice'', one whose edge lengths correspond to some multiple of the original lattice spacing. That is, arranging the $\tilde 5$-$\tilde 7$ cuts of the underlying lattice according to the super-lattice positions guarantees that the {\sl kirigami} elements are commensurate.

\begin{figure}
\centerline{\includegraphics[width=.4\textwidth]{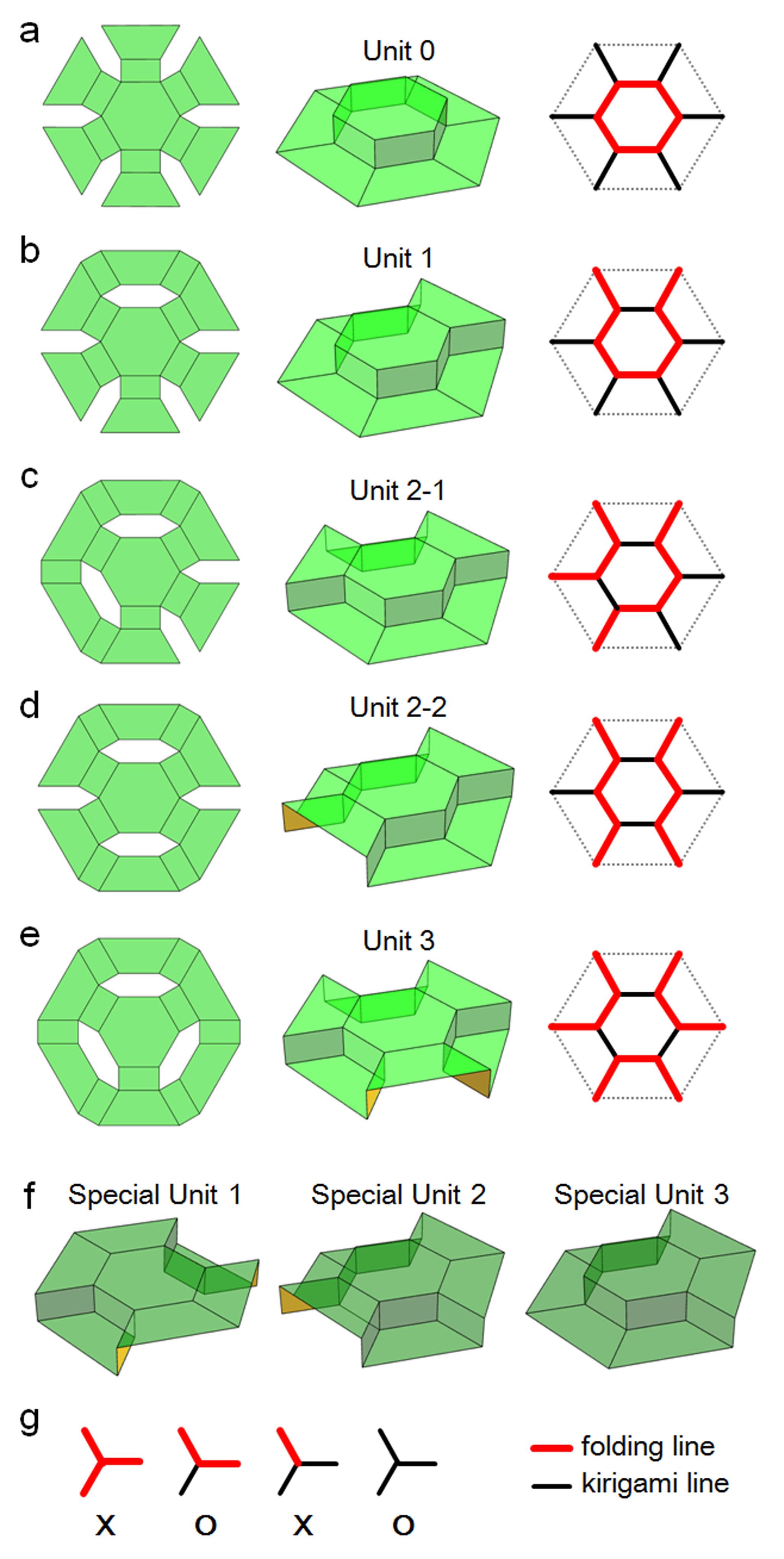}}
\caption{\label{fig:5_7_units}
(a) - (e) The basic building blocks of $\tilde 5$-$\tilde 7$ stepped surfaces. The left column shows the unfolded configuration, where the excised hexagons sit on a larger-scale honeycomb lattice. The middle column shows the folded configuration, and the right column shows a reduced representation suitable for easily designing target surfaces. (f) Folded configurations where the positive-climb paths of three dislocations converge. (g) Junction representation of the meeting of folding lines and cutting lines ({\sl i.e.} places where excised regions had their edges identified) in the reduced representation; only the junctions marked ``O'' represent allowed configurations.
}
\end{figure}

The left column of Fig. \ref{fig:5_7_units}(a)-(e) demonstrates the allowable ways (modulo rotations in $\mathbb{R}^3$) of decorating a given hexagon in the bulk of the super-lattice with $\tilde 5$-$\tilde 7$ elements (additional units are allowed at the boundaries of the sheet) while still maintaining the connectedness of the sheet. For instance, the structure labeled ``Unit 0'' is created with zero cuts and six folds around the central hexagon, which is accommodated by arranging six radial $\tilde 5$-$\tilde 7$ climb pair elements that each point at the corner of a hexagon on the super-lattice. The central column of Fig. \ref{fig:5_7_units} shows the folded configuration of each of these basic building blocks. In our nomenclature ``Unit $n$'' refers to a super-lattice hexagon with $n$ of its sides replaced with a $\tilde 5$-$\tilde 7$ climb pair element. The three structures in Fig. \ref{fig:5_7_units}(f) labeled ``special units" contain points at which three dislocations, whose Burgers vectors sum to zero, have positive-climb paths that converge together (as described in the Supplementary Material of \cite{Castle2014}). In real materials these special units may be less stable than the other units, but their use substantially simplifies the design of target structures.

The right column of Fig. \ref{fig:5_7_units} shows a simplified two-dimensional representation of each structure given by color-coding the edges of the hexagonal super-lattice as either folds or cuts (trivially done by comparing the relative elevation of the structure across a line). The allowable configurations can be determined by considering each vertex of the super-lattice, as shown in Fig. \ref{fig:5_7_units}(g): in order to avoid tearing and other unwanted structural deformations of the sheet, each vertex must have either zero or two folding lines incident upon it.

With this junction rule in hand it is straightforward to reverse engineer a two-dimensional map of the $\tilde 5$-$\tilde 7$ elements and fold lines needed to stepwise approximate a target surface. In the Supporting Information we demonstrate this by designing a {\sl kirigami} ziggurat, which we then realize in a simple experimental setting \cite{SupMat}. 

However, even aside from the restrictions imposed by the junction rules described above, these $\tilde 5$-$\tilde 7$ climb pair {\sl kirigami} designs share one of the fundamental limitations of the origami patterns: for every desired target surface an entirely new pattern of cuts has to be programmed into the sheet, and only then can the folds can be made. In many contexts we would prefer a pluripotent {\sl kirigami} blueprint. Similar in spirit to the universal hinge pattern for making generic polycubes \cite{hinge}, we desire a single arrangement of kirigami elements that can accommodate many different target structures, all accessed by entirely local sequences of fold reassignments ({\sl i.e.} changing a local set mountain folds to a valley folds, and {\sl vice versa}). In the next section we show that a triangular lattice of sixons can achieve precisely this goal.

\section{Deploying sixons for surface design}
\begin{figure}
\centerline{\includegraphics[width=.25\textwidth]{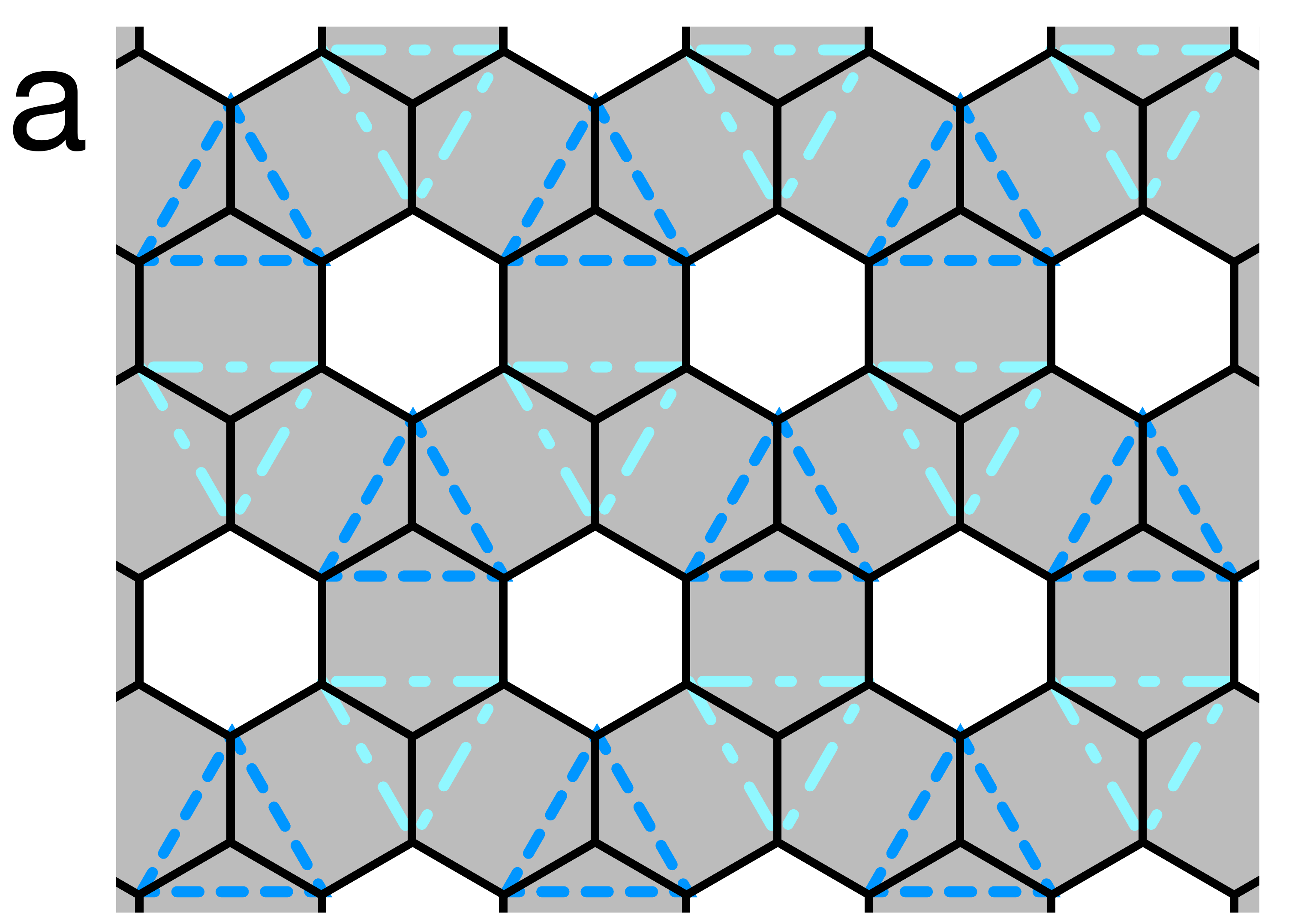}  \includegraphics[width=.25\textwidth]{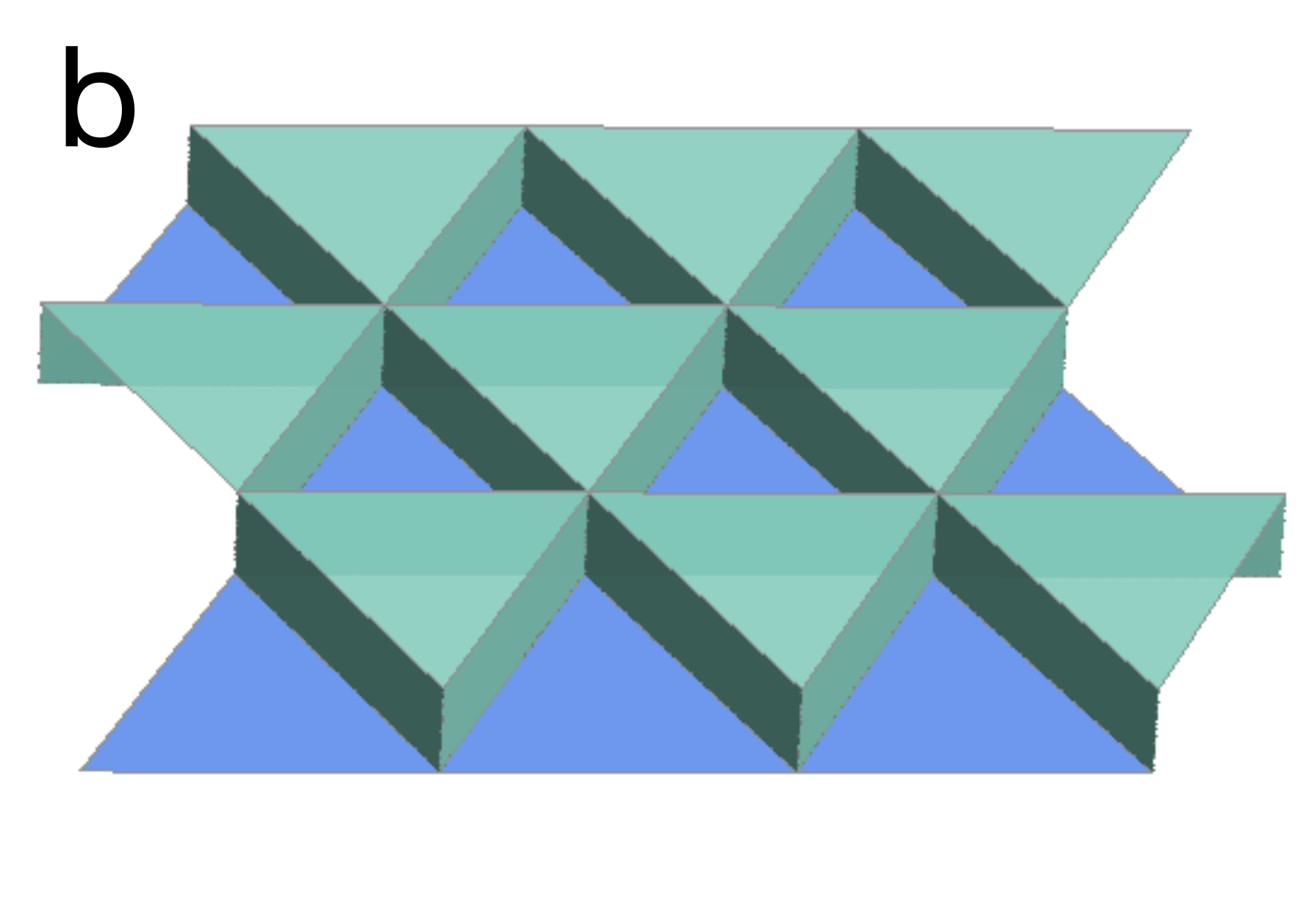}}\centerline{\includegraphics[width=.25\textwidth]{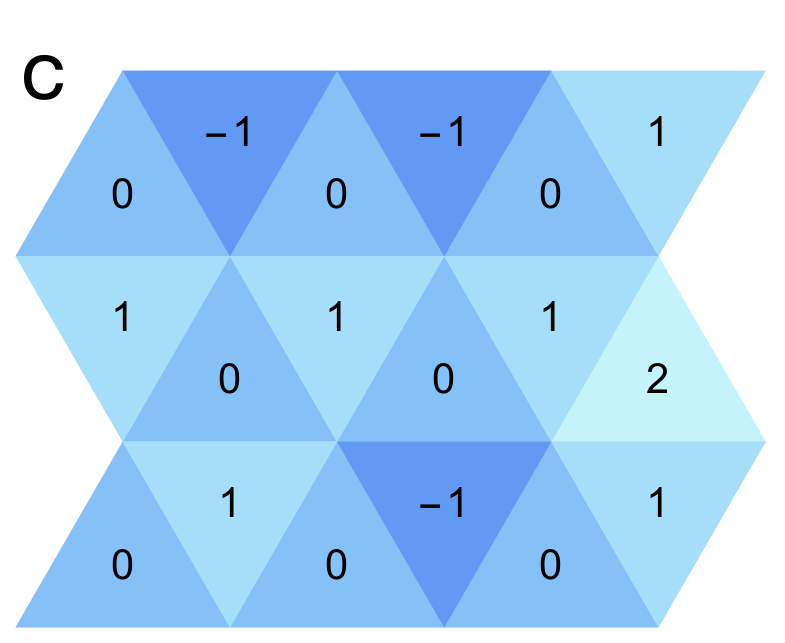}  \includegraphics[width=.27\textwidth]{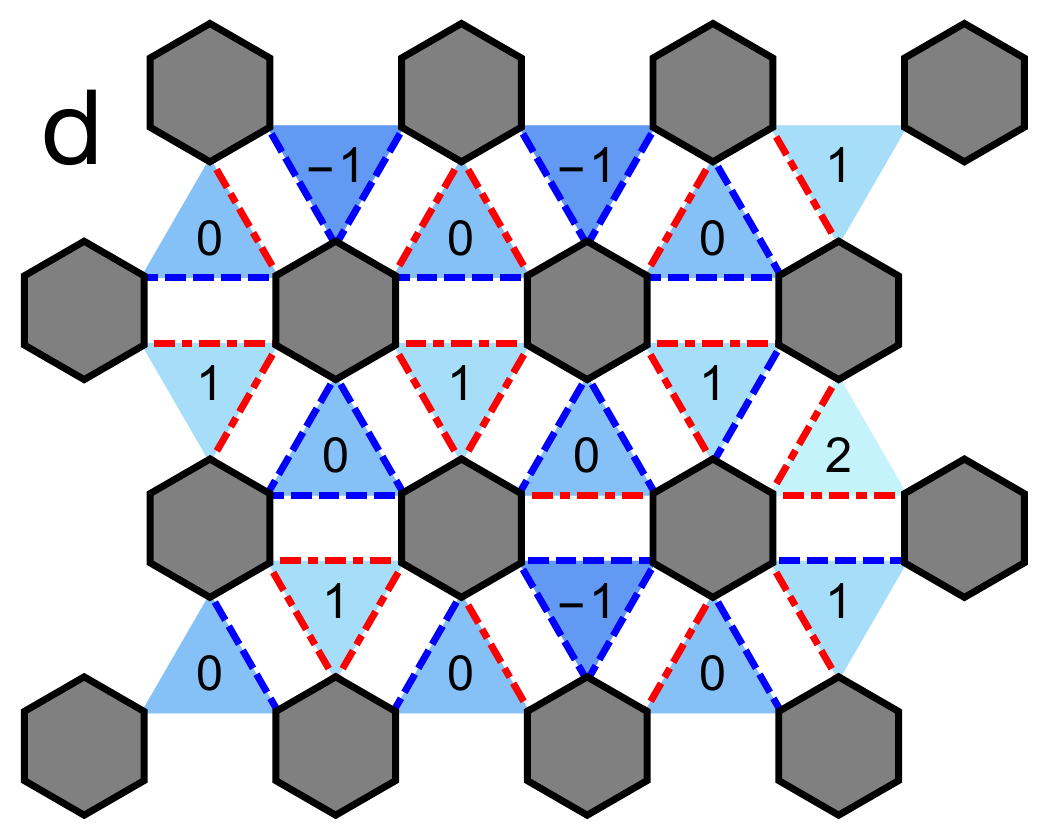}}
\caption{\label{fig:sixon_base}
(a) A triangular lattice of sixons together with lines for mountain (dot-dashed) and valley (dashed) folds to create the ground state configuration. (b) The ground state of the pattern in part (a), with sidewall heights reduced for visual clarity. (c) A reduced triangular representation for an arbitrary-height configuration of the lattice, where the numbers inside each triangle correspond to the height of that triangular plateau. Each plateau height differs by one from any plateau with which it shares an edge. (d) The cut-and-fold pattern corresponding to the height map in part (c), where the grey hexagons are the excised regions.
}
\end{figure}

As before, to make use of the sixon structure we must understand the allowed arrangements of sixons with respect to each other. Fortunately this is particularly straightforward: multiple sixons can be arranged in a triangular lattice. That is, the fold patterns required by the placement of multiple sixons can be made commensurate by choosing the excised hexagon centers to lie on a triangular lattice. The simplest implementation of this is shown in Fig. \ref{fig:sixon_base}(a). The folded state of this pattern, schematically shown in Fig. \ref{fig:sixon_base}(b), consists of triangular basins at height 0 separated by triangular plateaus at height 1 (in units of the length of the Burgers vector of the 2-4 cuts). By thinking of this structure in a gravitational potential we consider the configuration in Fig. \ref{fig:sixon_base}(b) to be the ground state of the triangular lattice of sixons, as it minimizes the sum of the relative plateau heights.

Local excitations of this configuration correspond to changing the plateau heights, for instance by interchanging the nested valley and mountain folds around a height-zero triangular basin, creating a height-two triangular plateau surrounded by height-one plateaus. If the sheet can bend and stretch as in bistable {\sl origami} \cite{Cohen2015}, transforming from the ground state to these excited states only requires opening and then re-closing the three excised hexagons that bound the relevant triangular plateau. Otherwise, to avoid tearing or bending while moving from one target to another, it is necessary to return to the completely flat state. The requirements that folds be commensurate and the sheet not stretch impose the only restriction on allowed excitations: the height of any triangle in the folded configuration must differ by one from the height of any triangle with which it shares an edge. One such excited state is shown in Fig. \ref{fig:sixon_base}(c); as shown in Fig. \ref{fig:sixon_base}(d), given a reduced representation of the heights it is trivial to assign the corresponding mountain and valley folds. 

\begin{figure*}
\centerline{\includegraphics[width=1.0\textwidth]{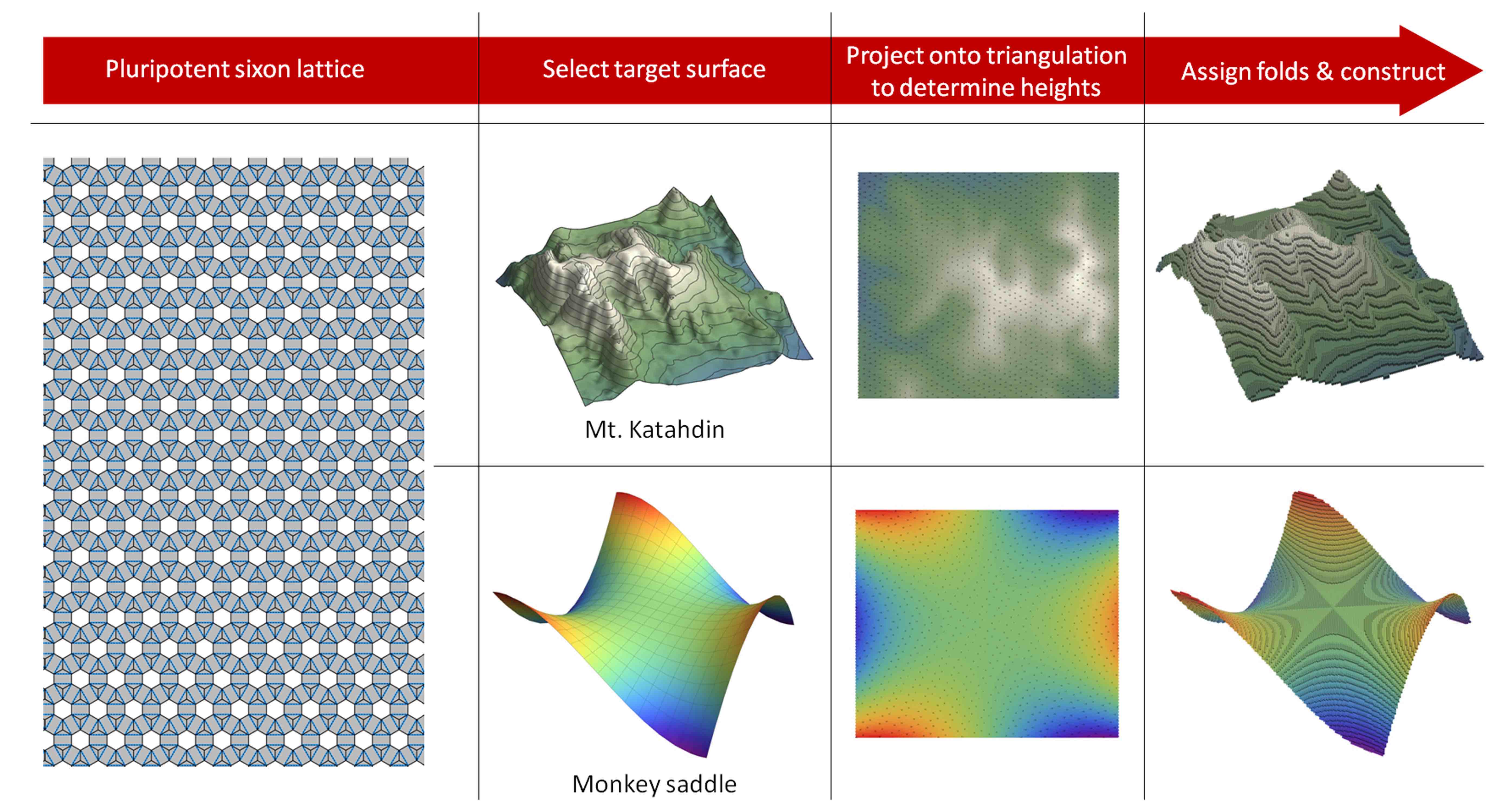}}
\caption{\label{fig:alg_design}
Illustration of the pluripotent design capability of the sixon lattice. Starting from the base configuration of a 151$\times$151 grid of triangular plateaus, a target surface is first selected (here a monkey saddle or Mt. Katahdin \cite{usgs}). The height of this target surface is projected onto the grid of triangular patches, and from this a local sequence of fold assignments is made to construct the final {\sl kirigami} structure.
}
\end{figure*}

The restriction on the excited states is quite modest, making it easy to design a cut-and-fold pattern based on a triangular lattice of sixons that approximately matches a target surface with a stepped surface of triangular plateaus. In this formulation, achievable target surfaces are limited only to have a maximum gradient set by the ratio of plateau height to plateau width. If the gradient of the target surface obeys this bound, finding the {\sl kirigami} solution to the inverse design problem is as straightforward as projecting the heights of a target surface down onto a triangular lattice, and then rounding the height assignment of each triangle to the nearest even or odd integer so that triangles sharing an edge will have heights that differ by exactly one. Indeed, if there is a known maximum gradient among the target surfaces into which a sixon array will be (re-)configured, the cut sizes determining the ratio of plateau height to width can be chosen to match this maximum gradient. From the triangular height map the pattern of mountain and valley folds immediately follows, as in Fig. \ref{fig:sixon_base}(d). We illustrate this simple design paradigm in Fig. \ref{fig:alg_design}, where the same triangular lattice of sixons can be used to approximate (to pick two arbitrary examples) both a monkey saddle and the topographic features at the northern terminus of the U.S. Appalachian Trail \cite{usgs}.

\section{Conclusion}
We have demonstrated that adding {\sl kirigami} cutting motifs to the folds of {\sl origami} leads to a powerful framework in which target structures can be algorithmically designed {\sl via} arrays of {\sl kirigami} elements. Subject to a gradient constraint, using a super-lattice simplifies the design for a static target surface, while triangular lattices of sixons form a versatile base pattern able to accommodate fold patterns for any stepped triangulated target structure. If the material allows for dynamic changing of fold type then the sixon lattice can be reconfigured between arbitrary surface configurations.  Current studies involve designing pluripotent {\sl kirigami} templates that lift the gradient limitation as well as other geometric and topological limits. We note that our choice of triangular plateaus and sixons was a powerful but not entirely unique choice with which to engineer a pluripotent sheet of kirigami elements. We show in the Supporting Information that using square plateaus leads to a similar (but more restrictive) design paradigm \cite{SupMat}. 

Of particular importance, we note that our {\sl kirigami} designs are superior to many {\sl origami} designs for comparable results, especially in relation to self-assembly. For example, our experimental realization of a ziggurat-style pryamid \cite{SupMat} -- actuated by heat-shrink tape placed across the short axis of the excised hexagons -- makes it clear that in our {\sl kirigami} constructions the fold process is very robust. That is, in contrast with the {\sl origami} methods discussed above, detailed control over the folds is not necessary to correctly form the target structure. Also, unlike some {\sl origami} designs, this {\sl kirigami}-based approach to approximating surfaces with sixons maintains a constant complexity of fold pattern per unit surface area. The underlying structure is always a triangular lattice of sixons, and different surfaces differ in their folding template only by switching mountain and valley folds as needed. 

Because the folds connecting sixons are coupled, a triangular lattice of sixons has more fold lines than degrees of freedom. This attribute could be exploited to create truly multi-potent kirigami sheets by partitioning the folds into non-overlapping sets and programming each of those sets to fold in response to different stimuli. We schematically illustrate such a ``duo-potent'' sheet in the Supporting Information \cite{SupMat}, programming half of the fold lines to form a half-cylinder when activated, while the other half of the fold lines transform the sheet into a Mexican hat potential. This raises the exciting prospect that, with dynamical control over the fold type, a single lattice of sixons could serve as the base for, {\sl e.g.} microfluidic devices with dynamically changeable channels. Work on implementing such a dynamically reconfigurable {\sl kirigami} surface is currently underway. We also anticipate that exploring the collective elasticity of these {\sl kirigami} sheets will provide further insight into two-dimensional mechanical metamaterials \cite{Paulose}.

\begin{acknowledgments}
The authors acknowledge support from NSF EFRI-ODISSEI Grant EFRI 13-31583.  D.M.S. was supported by the Advanced Materials Fellowship from the American Philosophical Society. This work was partially supported by a Simons Investigator grant from the Simons Foundation to R.D.K.
\end{acknowledgments}

\clearpage


\section{Supporting information}

\setcounter{table}{0}
\renewcommand\thefigure{S\arabic{figure}}

In the first section of this supporting information we comment on actively controlling subsets of folds to make a duopotent sixon sheet  In the second section we provide additional details of our $\tilde 5$-$\tilde 7$ {\sl kirigami}  design protocol, including an experimental demonstration in which a {\sl kirigami} pattern self-folds into a ziggurat. In the third section we discuss how our design rules behave on square lattices.

\section{Pluripotent sixons}
\begin{figure}[!hb] 
\centerline{\includegraphics[width=0.61\linewidth]{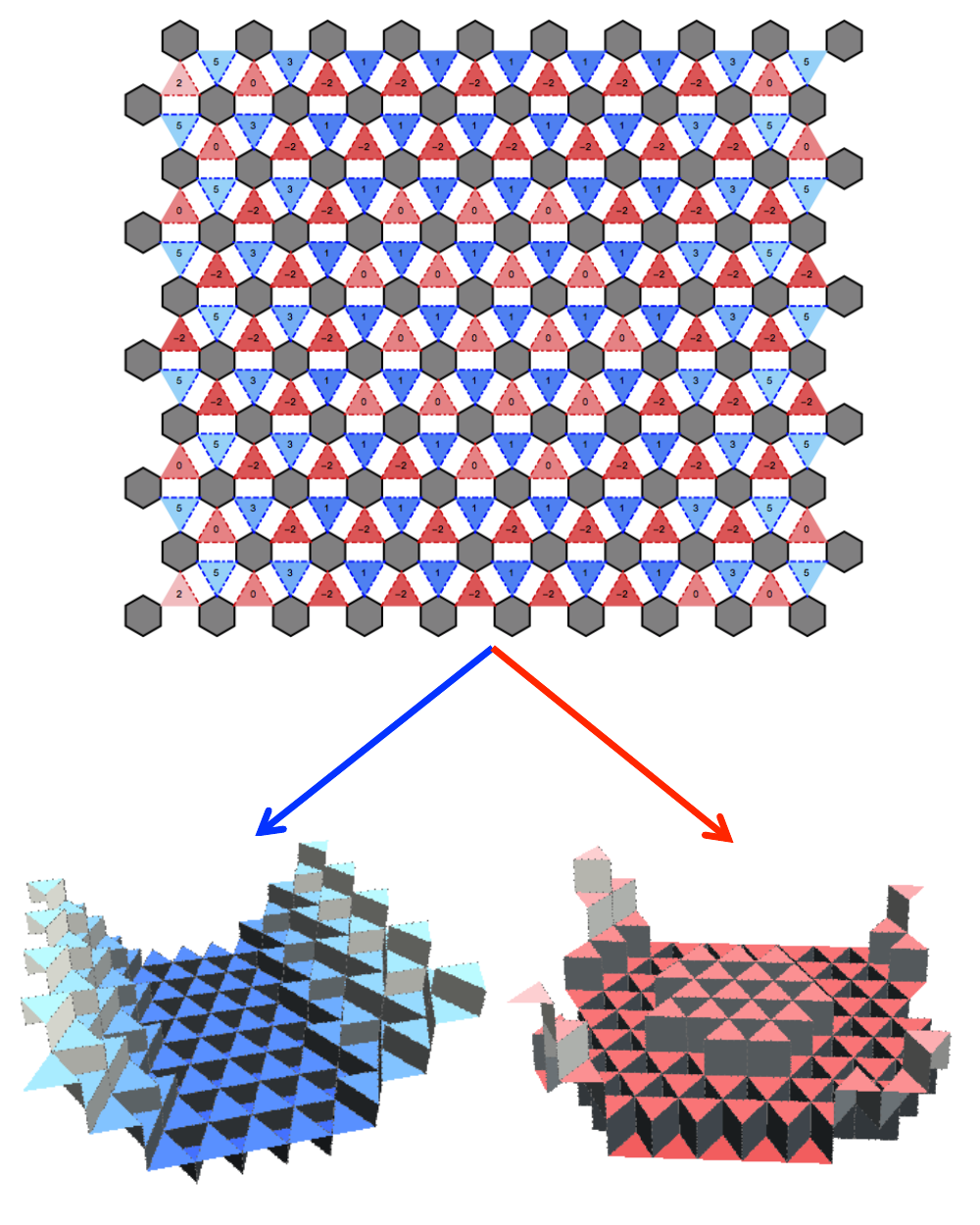}}
\caption{\label{fig:SI_duopotent}
(Top) A duopotent lattice of sixons in the flat state, where grey hexagons denote the excised regions of the paper. Folding the red lines (surrounding the red, upward-facing triangles) leads to a mexican-hat potential, whereas folding the blue lines (surrounding the blue, downward-facing triangles) leads to a half-cylinder. Dashed lines indicate valley folds and dash-dotted lines indicate mountain folds. (Bottom) The two folded-state configurations of the duopotent sheet.}
\end{figure} 

As described in the main text, a triangular lattice of sixons has the potential to fold to any target conformation that satisfies the maximum-gradient constraint. Once the hexagons are excised it is the fold lines between these hexagons that will then determine the final conformation. If the folds could be dynamically reprogrammed -- certainly possible using electrically actuated hinges, and perhaps with other implementations as well -- then the surface could transition between different configurations of the dynamically pluripotent sheet, using the flat configuration as an intermediate state. 

In practice, depending on the method used it may be difficult to implement complete dynamic control over every fold in a lattice of sixons. Nevertheless, there are more fold lines than degrees of freedom in a sixon lattice, and this fact can be exploited to create multi-potent {\sl kirigami} sheets. The idea is to pre-program subsets of fold lines to bend in particular directions; given that there are more folds than degrees of freedom one can completely specify multiple target surfaces by patterning multiple such subsets to actuate in response to different external stimuli. This works even if one is limited to actuating each fold line only in response to one of the external stimuli, {\sl i.e.} using disjoint subsets of folds to control target structures.

A robust implementation of this is schematically shown in the ``duo-potent'' sheet in Fig. \ref{fig:SI_duopotent}, which exploits the fact that around every triangular plateau are two concentric folding-line triangles. Controlling, {\sl e.g.}, all of the fold lines surrounding the ``upward facing'' triangles guarantees that the complementary folds automatically complete as the surface moves from the flat state to the target configuration with all of the holes closed. In this way programming the upward-facing triangles specifies one target surface, while similarly programming the downward-facing triangles specifies a second independent target surface. In Fig. \ref{fig:SI_duopotent} we have illustrated this idea by programming the folds of one set of triangles to produce a half-cylinder, while the other triangles define the beginnings of a ``mexican-hat potential''. It is our hope that such combined configurations can be used by experimentalists to verify the utility of the sixon lattice as a reconfigurable medium by using different stimuli to activate each set of folds.

\section{$\tilde 5$-$\tilde 7$ {\sl kirigami} design}
\subsection{Design rules}
In the main text we showed that the basic units for this method of {\sl kirigami} design are $\tilde 5$-$\tilde 7$ climb-pair elements placed on the honeycomb super-lattice. Designing the cuts and folds to achieve a target surface is most easily done by projecting the height of the target surface onto the faces of the super-lattice: these faces become hexagonal plateaus in the coarse-grained ``pixellated'' {\sl kirigami} version of the surface. The junction rules, which state that there should be either zero or two folding lines (and thus either one or three cut lines) incident upon each vertex of the honeycomb super-lattice, greatly simplify the task of arranging the $\tilde 5$-$\tilde 7$ on the super-lattice. Additionally, to maintain the connectedness of the surface no hexagon should be surrounded by six cutting lines. 

\begin{figure}
\centerline{\includegraphics[width=\linewidth]{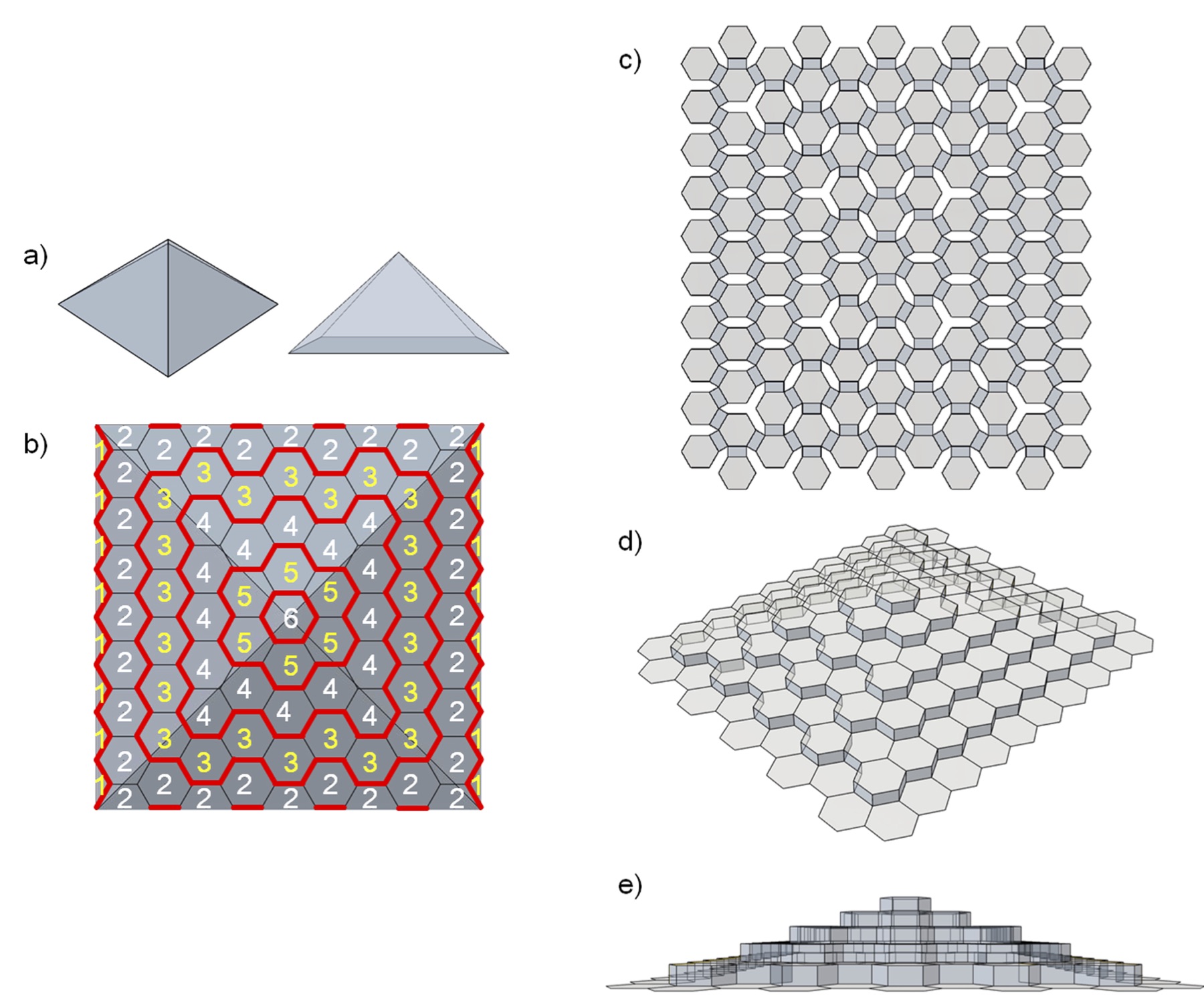}}
\caption{\label{fig:SI_pyramid}
$\tilde 5$-$\tilde 7$ climb pair design of a ziggurat. (a) The target surface. (b) Projecting the integer heights of the target surface down onto a plane decorated with the honeycomb super-lattice. As in the main text, red lines indicate folds and black lines indicate cuts, while the numbers indicate the height of the labeled hexagonal plateau. (c) Cut-and-fold pattern corresponding to the projected pattern. (d)-(e) Computer renderings of the final, folded state.}
\end{figure}

Within this framework, the height difference between adjacent hexagons cannot be larger than one step (the height of the vertical side walls), and each hexagon must have at least one adjacent hexagon at a different height. These constraints set the ultimate resolution that we are able to achieve with these designs. The junction rules also impose some minor non-local restrictions on the allowed combination of  {\sl kirigami} units, and depending on the complexity of the target surface the initial projection of heights may have to be modified to satisfy these constraints. Nevertheless, target surfaces that do not violate the above restrictions can be designed by joining basic {\sl kirigami} units together. As a proof of concept, we choose a pyramidal target surface. As illustrated in Fig. \ref{fig:SI_pyramid}, we first project the height of the pyramidal surface onto the super-lattice. The junction rules illustrated by the height projection are then easily turned into a cut-and-fold pattern, and the final state of this pattern is then rendered.

\subsection{Experimental realization}
\begin{figure}
\centerline{\includegraphics[width=0.5\linewidth]{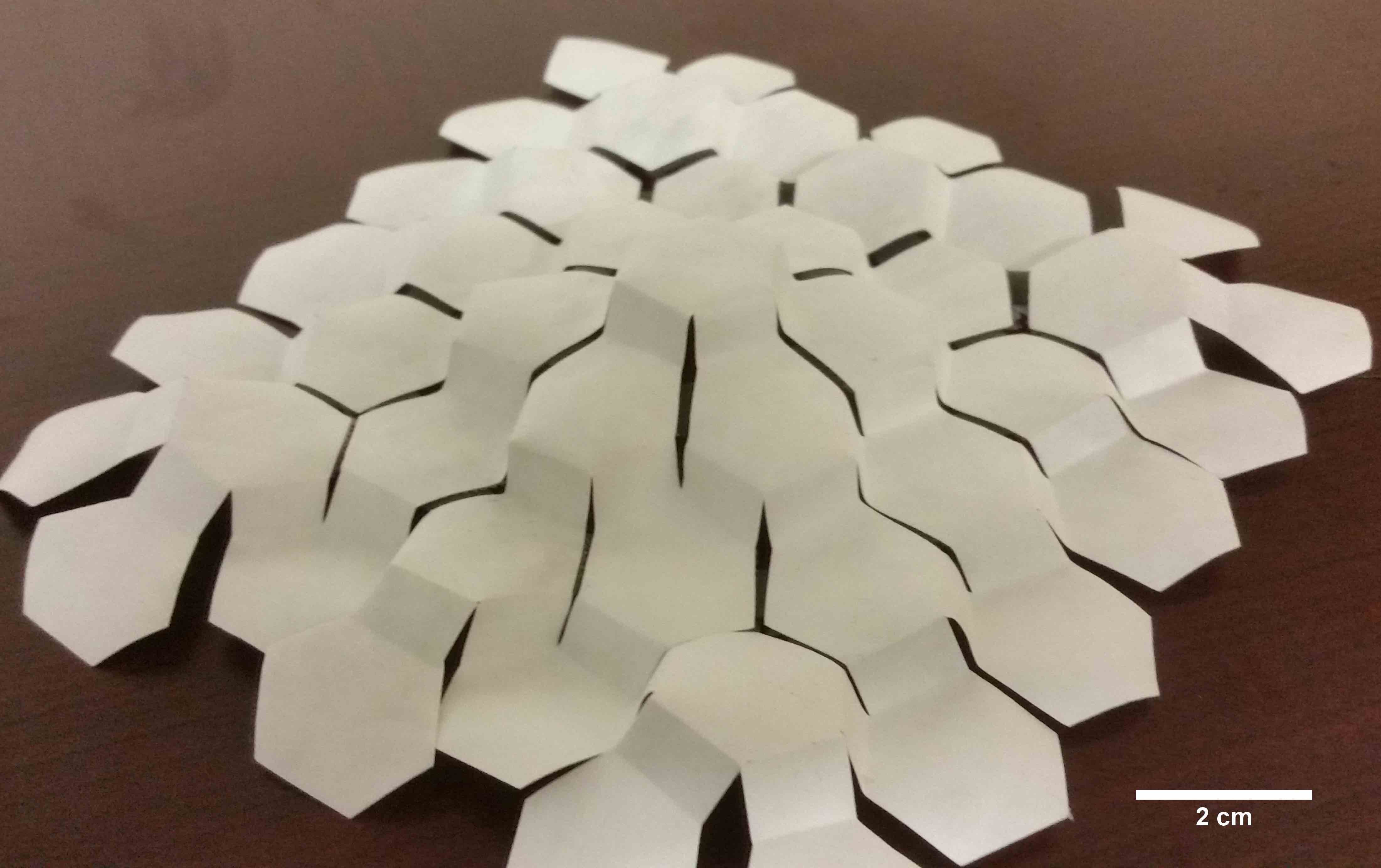}}
\caption{\label{fig:SI_expt} Ziggurat {\sl kirigami} constructed from Tyvek and polyolefin (video of this structure self-folding available at https://dl.dropboxusercontent.com/u/28290913/zigguratkirigami.mp4).}
\end{figure}

To complement our schematic designs in the main text and above, we demonstrate a particularly simple experimental setting in which our {\sl kirigami} design principles can be implemented. We used Tyvek  (nonwoven, Spunbonded Olefin, Type 10) as our two-dimensional material because it is stronger and more tear-resistant than standard paper.  We cut the pattern in Fig. \ref{fig:SI_pyramid}(c) onto the Tyvek, pre-creased the folds, and then bonded heat-shrinkable polyolefin (SPC Technology) under the hexagonal cuts by hot pressing at 120$^\circ$C for 20 minutes.  To prevent shrinking during bonding, the assembly was kept pressed until cooled to room temperature. Subsequent (unpressed) baking at 95$^\circ$C for one minute curled the polyolefin strips and folded the Tyvek into the target structure shown Fig. \ref{fig:SI_expt}. With the folds precreased this method led to sharper steps in the folded state compared to applying the polyolefin directly to the fold lines. This autonomous self-assembly of our three-dimensional target structures demonstrates the robustness of our {\sl kirigami} rules.

\section{Square cut design}

\begin{figure}[!hb] 
\centerline{\includegraphics[width=0.7\linewidth]{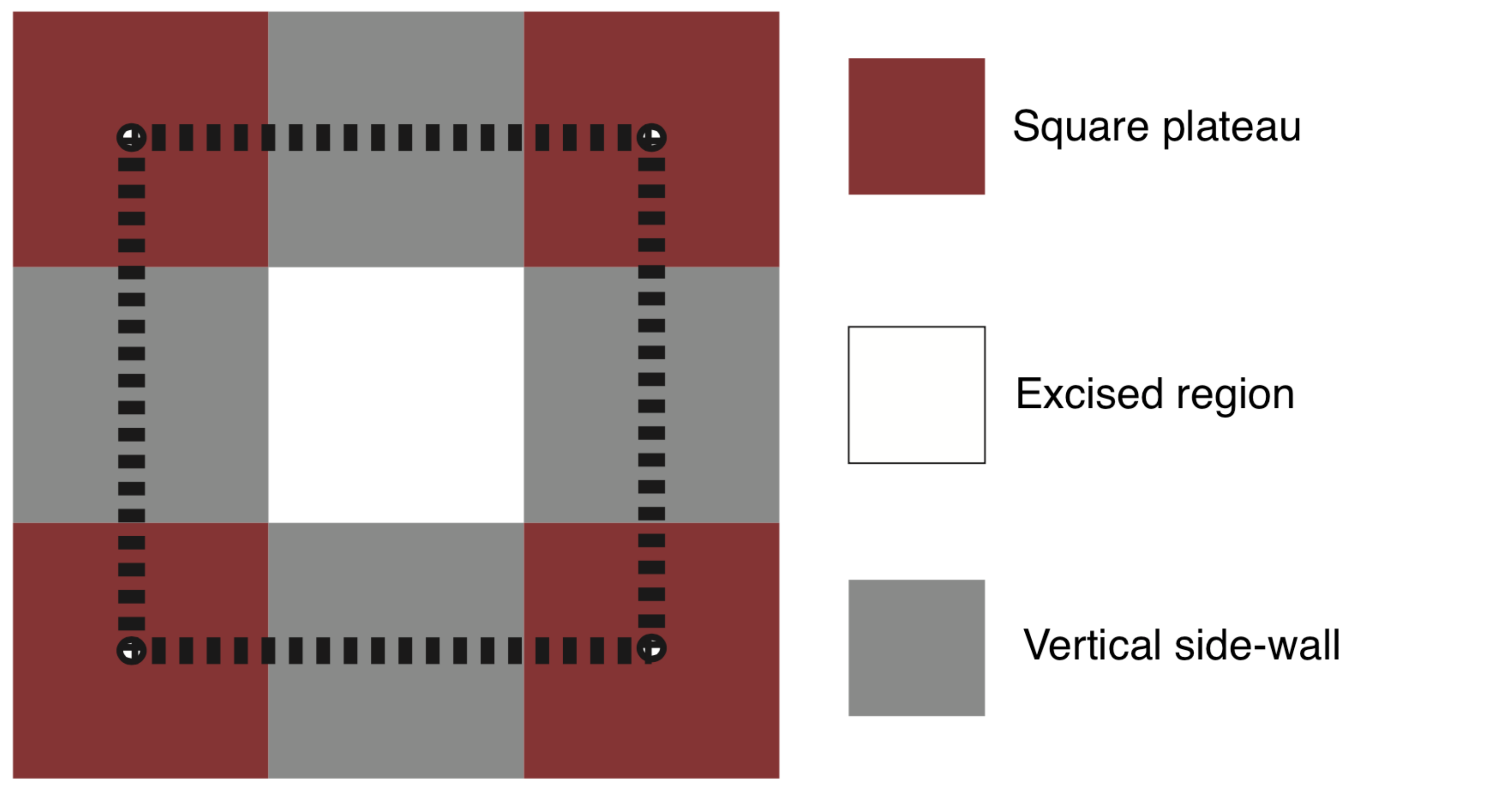}}
\caption{\label{fig:SI_square_unit_base}
The base unit of square-lattice {\sl kirigami}. The part of the figure on the left enclosed by dotted lines is used to tile the plane, and the colors encode the ultimate fate of each square patch in the folded configuration.}
\end{figure}

Our choice of a honeycomb lattice in the main text and above on which to base our {\sl kirigami} designs is far from unique. Different lattices have different concomitant angles and will thus generate {\sl kirigami} surfaces with differently shaped tiles. One obvious alternative to the hexagonal lattice is the square lattice, and modifying the sixon technique in the natural way generates a {\sl kirigami} surface with square tiles. Fig. \ref{fig:SI_square_unit_base} shows the basic unit of square-lattice {\sl kirigami}: the portion enclosed by dotted lines tiles space, the white central squares are excised, the red squares at the corners become the square plateaus, and the gray squares become the vertical side walls. In this formulation the sidewall height is set to be the same as the plateau width; to independently vary the plateau size and sidewall heights the plateaus and excised squares remain square but change in relative size, while the sidewalls become rectangular. 

\begin{figure}[!hb] 
\centerline{\includegraphics[width=0.7\linewidth]{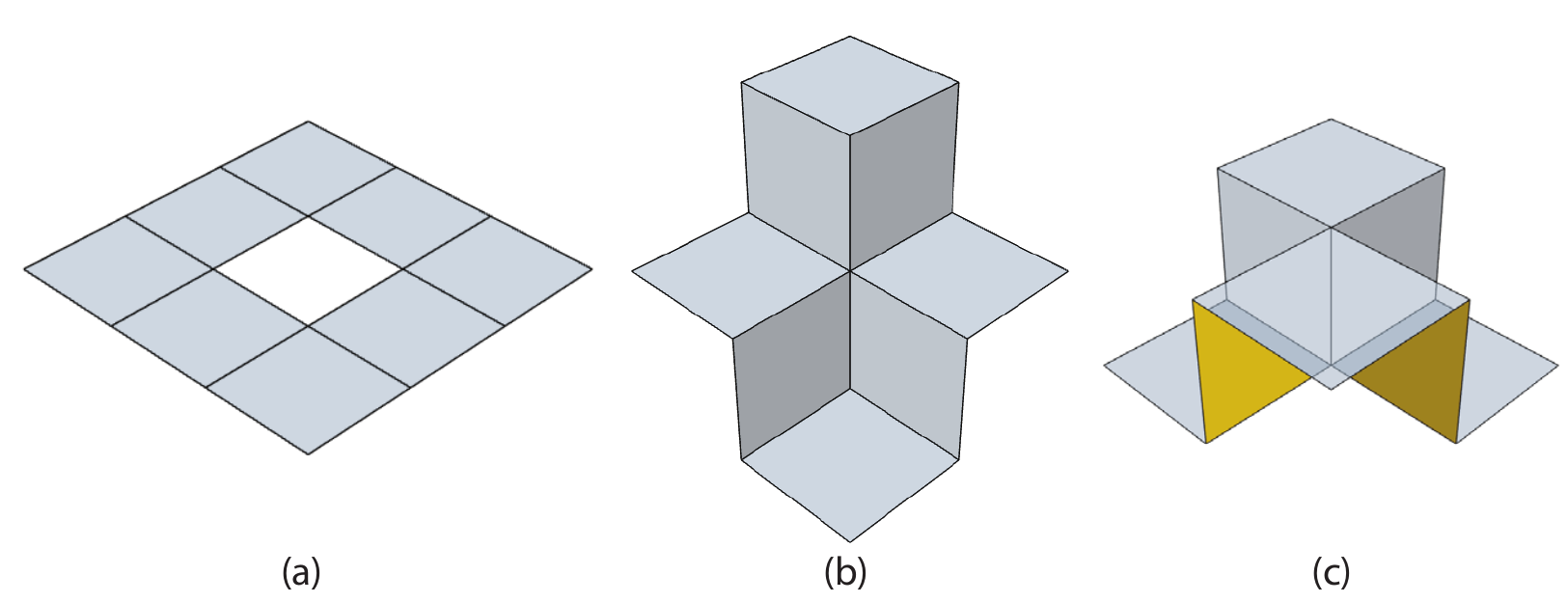}}
\caption{\label{fig:SI_square_iso}
Construction and configurations of the fundamental square-lattice {\sl kirigami} element (a) The square-lattice defect in its unfolded state. (b)-(c) Two folded configurations of this element, which capture the allowed states up to rotations.}
\end{figure}

\begin{figure}[!hb] 
\centerline{\includegraphics[width=0.7\linewidth]{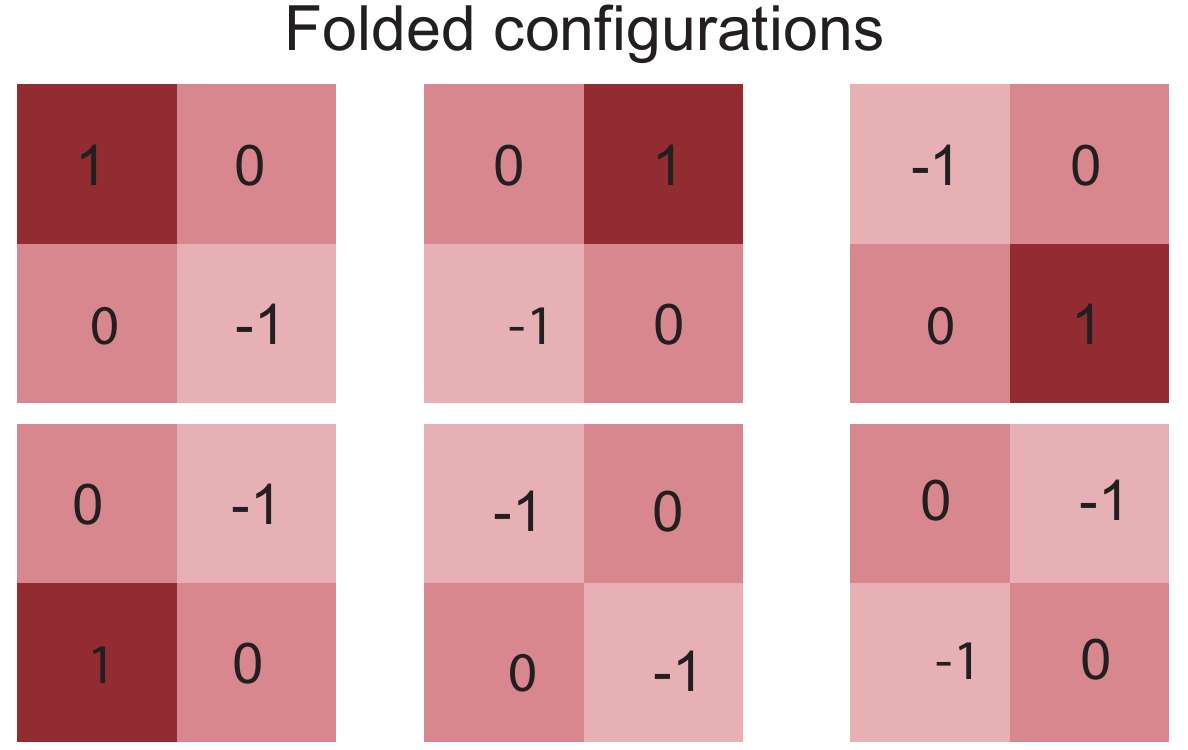}}
\caption{\label{fig:SI_square_units}
Reduced representation of the square-lattice {\sl kirigami} building blocks. Target configurations can be formed by overlapping these 6 units (up to addition of a constant integer to each of the four square heights)}
\end{figure}

Fig. \ref{fig:SI_square_iso} shows the two symmetrically distinct configurations formed by identifying the edges of the of the excised square in different ways. A desired target structure can then be designed by stitching these units together, just as in the $\tilde 5$-$\tilde 7$ {\sl kirigami} design paradigm. This again allows a stepped-surface representation of target surfaces, which is programmed by projecting the nearest odd or even (whichever is appropriate) integer heights of the target surface onto the square lattice. Allowed folded-state structures are shown in Fig. \ref{fig:SI_square_units}. Although some of the six structures depicted are related to each other via rotations, their enumeration provides a convenient check on the globally allowed configurations. In particular, for a given assignement of heights to be globally allowed, every $2 \times 2$ subset of the height map must be one of the configurations shown in Fig. \ref{fig:SI_square_units}  (up to adding a constant integer to each of the four square heights). This criteria, rather than the junction rules in the $\tilde 5$-$\tilde 7$ {\sl kirigami}, provides a way of capturing  the non-local restrictions the square-cut designs have.

\end{document}